\documentclass[aps,pre,twocolumn,superscriptaddress,showpacs]{revtex4}
\usepackage{graphicx}
\usepackage{hyperref}
\usepackage{natbib}
\usepackage{bm}

\begin{document}

\title{An efficient scheme for numerical simulations of the 
spin-bath decoherence}
\author{V. V. Dobrovitski}
\affiliation{Ames Laboratory, Iowa State University, Ames IA 50011, USA}
\author{H. A. De Raedt}
\affiliation{Department of Applied Physics/Computational Physics, Materials
  Science Centre, University of Groningen, Nijenborgh 4, NL-9747 AG
  Groningen, The Netherlands}

\begin{abstract}
We demonstrate that the Chebyshev expansion method is
a very efficient numerical tool for studying spin-bath decoherence
of quantum systems.
We consider two typical problems arising in studying 
decoherence of quantum systems consisting of few coupled
spins: (i) determining the
pointer states of the system, and (ii) determining
the temporal decay of quantum oscillations. As our
results demonstrate, for determining the pointer states,
the Chebyshev-based
scheme% has been faster by a factor of more than 8,
is at least a factor of 8 faster
than existing % fast
algorithms based on the Suzuki-Trotter decomposition.
For the problems of second type, the Chebyshev-based
approach has been 3--4 times faster than the Suzuki-Trotter-based
schemes. This conclusion holds qualitatively for a wide spectrum of
systems, with different spin baths and different Hamiltonians.
\end{abstract}

\date{\today}
\pacs{75.10.Jm, 02.60.Cb, 75.10.Nr, 03.65.Yz}

\maketitle

\section{Introduction}
Recently, a great deal of attention has been devoted 
to the study of quantum computation \cite{1,2}.
For many physical systems, basic quantum operations needed for
implementation of quantum gates have been demonstrated.
To be practical, a quantum computer
should contain a large number of qubits (some estimates
give up to 10$^6$ qubits \cite{8}), and be able to perform many
hundreds of quantum gate operations. However, 
these requirements are not easy to satisfy in experiments.
A real two-state quantum
system is different from the ideal qubit. The system interacts with 
its environment, and this leads to
a loss of phase relations between different states of the
quantum computer (decoherence) \cite{9,9a,10,11}, causing rapid %generation
accumulation of errors. Detailed theoretical understanding of
the decoherence process is needed to prevent this.

More generally, decoherence is an interesting many-body quantum phenomenon
which is fundamental for many areas of quantum mechanics,
quantum measurement theory, etc \cite{9,9a}. It also
plays an important role in solid state systems, and might
suppress quantum tunneling of defects in crystals 
\cite{11}, spin tunneling in magnetic molecules and
nanoparticles \cite{qtm94,keiji}, or destroy Kondo effect
in a dissipationless manner \cite{kondo}. 
I.e., decoherence in many physical systems can have 
experimentally detectable (and sometimes considerable)
consequences, and extensive theoretical studies of 
decoherence are needed to understand behavior of
these systems.

Formally speaking, decoherence is a dynamical development
of quantum correlations (entanglement) between the
central system and its environment. Let us assume that initially the
central system is in the state $|\psi_0\rangle$ and the
environment is in the state $|\chi_0\rangle$, so that
the state of the compound system (central system plus bath)
is $|\Psi(t=0)\rangle =|\psi_0\rangle\otimes|\chi_0\rangle$. 
In the course of dynamical evolution, the direct product 
structure of the state $|\Psi(t)\rangle$ is no longer
conserved. If we need to study only the properties of
the central system, we can consider the reduced density 
matrix of the central system, i.e.\ the matrix
$\rho_S(t) = \mathop{\rm Tr}_B |\Psi(t)\rangle\langle\Psi(t)|$,
where $\mathop{\rm Tr}_B$ means tracing over the environmental
degrees of freedom. Initially, 
$\rho_S(0)=|\psi_0\rangle\langle\psi_0|$, the system
is in pure state, and its density matrix is a projector,
i.e.\ $\rho^2_S(0)=\rho_S(0)$. At $t>0$, this property
is lost, and the system appears in a %the
mixed state.
It has been shown that, even for relatively small
integrable and non-integrable systems,
the mixing is sufficient for the time-averaged, quantum dynamical
properties of the subsystem to
agree with their statistical mechanics values~\cite{keiji2}.
Diagonalizing the density matrix $\rho_S$, we can find
the (instantaneous) states of the system $|q_i(t)\rangle$
and (instantaneous) occupation numbers of these states $w_i(t)$.
It is generally assumed (and is true for all cases we know) that
in ``regular'' situations, the states $|q_i(t)\rangle$ quickly relax
to some limiting states $|p_i\rangle$, called ``pointer states''.
This process (decoherence) is, in most cases, much faster
than the relaxation of the occupation numbers $w_i(t)$ to
their limiting values (which correspond to thermal equilibrium
of the system with the bath).

The theoretical description of decoherence, i.e.\ a
description of the evolution
of the central system from its initial pure state $\psi_0$ to
the final mixed state, and finding the final
pointer states $|p_i\rangle$, is a very difficult
problem of quantum many-body theory. Some simple models
can be solved analytically, for some more complex models
different approximations can be employed, such as 
the Markov approximation for
the bath, which %presumes
assumes that the memory effects in
the bath dynamics are negligible. A special case of
environment consisting 
of uncoupled oscillators, so-called ``boson bath'', is
also rather well understood theoretically. But,
although the model
of boson bath is applicable for description of 
a large number of possible types of environments 
(phonons, photons, conduction electrons, etc.) \cite{11},
it is not universal.

A particularly important case where the boson bath
description is inapplicable is the decoherence caused by
an enviroment made of spins, e.g.\ nuclear spins, or
impurity spins (so called ``spin bath'' environment). 
Similarly, decoherence caused by some other types of
quantum two-level systems can be described in terms of
the spin bath. Analytical studies of the spin-bath 
decoherence are difficult, and the spin-bath decoherence
of many-body systems is practically unexplored yet.
In this situation, numerical modeling of spin-bath 
decoherence becomes an invaluable research tool.

The most direct approach to study spin-bath decoherence
is to compute the dynamical evolution of the
whole compound system by directly solving the time-dependent
Schr\"odinger equation of the model system.
Even for a modest amount of spins, say 20, such calculations
require considerable computational resources, in particular
because to study decoherence we have to follow the dynamical
evolution of the system over substantial periods of time.
Therefore it is worthwhile to explore ways to significantly
improve the efficiency of these simulations.

In this paper, we apply the Chebyshev's
expansion method 
to simulate models for the spin-bath decoherence.
This method has been widely applied before 
\cite{TAL-EZER0,TAL-EZER,LEFOR,kosloff1,Iitaka01,SILVER,LOH}
to study dynamics of large quantum systems, but, to
our knowledge, has never been used for simulations of
systems made of large number of coupled quantum spins.
We show that for realistic problems and typical values
of parameters this method is a very efficient tool,
giving significant increase in the
simulations speed, sometimes up to a factor of eight,
in comparison with the algorithms \cite{hdr,hdr1} based on Suzuki-Trotter
decompositions \cite{SUZUKI1}.
We illustrate this point by test examples
that we have encountered in our previous
studies of the dynamics of the spin-bath decoherence.
%
% new insert
We also briefly discuss two other approaches, the 
short iterative Lanczos (SIL) method \cite{LEFOR,sil,jaklic}
and the multi-configurational time-dependent Hartree (MCTDH)
\cite{mctdh1,mctdh2,mctdh3} method, which are known to 
demonstrate very good performance in many
problems of quantum chemistry.
% end of new insert
%

The remainder of the paper is organized as follows. In
Section \ref{sec2}, we describe the model and the
approaches used for the decoherence simulations.
In Section \ref{sec3}, we describe the specific details
of application of the Chebyshev's expansion method
to the spin-bath decoherence simulations. In Section
\ref{sec4}, we present the results of our test 
simulations. A brief summary is given in the Section
\ref{secsum}.

\section{Simulations of the spin-bath decoherence: the model
and numerical approaches}
\label{sec2}

We focus on decoherence in quantum systems of several coupled spins.
This type of quantum systems is of particular interest for quantum computations,
since a qubit can be represented as a quantum spin 1/2, and qubit-based quantum computation
is, in fact, a controlled dynamics of the system made of many spins 1/2.
Such systems are also of primary interest for studying many solid state
problems, since an electron is a particle with the spin 1/2,
and its orbital degrees of freedom are often irrelevant.
Thus, a system made of several coupled spins 1/2
is a good model for investigating a large class of
important problems both in quantum computing, and in
solid state theory. The approach described below can be
easily extended to arbitrary spin values, but 
discussion of simulations with arbitrary spins is beyond
the scope of this paper.

We consider the following class of models.
There is a central system made of $M$ coupled spins
${\bf S}_m$ ($S_m=1/2$, $m=1\dots M$). The spins 
${\bf S}_m$ interact with a bath consisting of $N$ environmental
spins ${\bf I}_n$ ($I_n=1/2$, $n=1\dots N$). 
The Hamiltonian governing behavior of the whole ``compound''
system (central spins ${\bf S}_m$ plus the bath spins
${\bf I}_n$) is
\begin{equation}
{\cal H} = {\cal H}_0 + {\cal V} = {\cal H}_S + {\cal H}_B + {\cal V},
\label{genham}
\end{equation}
where ${\cal H}_S$ and ${\cal H}_B$ are the ``bare'' Hamiltonians
of the central system and the bath, correspondingly, and
${\cal V}$ is the system-bath interaction. Below, we
present simulation results for the following general form the Hamiltonians:
\begin{eqnarray}
\nonumber
{\cal H}_S &=& \sum_{\langle m,m'\rangle} \sum_{\alpha=x,y,z}
  J^{\alpha}_{mm'} 
  S^{\alpha}_m S^{\alpha}_{m'} + \sum_m \sum_{\alpha=x,y,z}
  H^\alpha_m S^{\alpha}_m,\\
  \nonumber
{\cal H}_B &=& \sum_{\langle n,n'\rangle} \sum_{\alpha=x,y,z}
  \Gamma^{\alpha}_{nn'} 
  I^{\alpha}_n I^{\alpha}_{n'} + \sum_n \sum_{\alpha=x,y,z}
  H^\alpha_n I^{\alpha}_n,\\
{\cal V} &=& \sum_{\langle m,n\rangle} \sum_{\alpha=x,y,z}
  A^{\alpha}_{mn} S^{\alpha}_m I^{\alpha}_{n}.
\label{specham}
\end{eqnarray}
We assume
that the Hamiltonian ${\cal H}$ does not explicitly depend 
on time, i.e.\ all exchange interaction constants $J$, $\Gamma$,
and $A$, and all external magnetic fields ${\bf H}$ are constant in time.
Although this makes impossible to model the
time-dependent quantum-gate operation, the investigation of the
fundamental properties of spin-bath decoherence is
not seriously affected by this requirement.
The dynamics of the model (\ref{genham}) is already too complex to be
studied analytically, and for general ${\cal H}$,
when no {\it a priori\/} knowledge is available,
the only option is to solve the time-dependent
Schr\"odinger equation of the whole compound system numerically.
I.e., we choose some basis states for the Hilbert space of
the compound system (the simplest choice is the direct
product of the states $|\uparrow\rangle$ and $|\downarrow\rangle$
for each spin ${\bf S}_m$, ${\bf I}_n$). We 
represent an initial state of the compound system
$\Psi_0$ as a vector in this basis set, and the Hamiltonian
${\cal H}$ is represented as a matrix, so that the 
Schr\"odinger equation
\begin{equation}
i\, d\Psi(t) / dt = {\cal H}\Psi(t)
\label{tdse}
\end{equation}
is a system of first-order ordinary differential equations
with the initial condition $\Psi(t=0)=\Psi_0$.

The length of the vector $\Psi$ is $2^{M+N}$; for typical
values $M=2$ and $N=20$, an exact solution of about $2\cdot 10^6$
differential equations becomes a serious task. Moreover,
the interaction between the central spins is often much bigger
than the coupling with environment or coupling between the
bath spins, so that the system (\ref{tdse}) is often stiff.
Simple methods, e.g.\ predictor-corrector schemes, perform rather
poorly in this case, and very small integration steps are
needed to obtain a reliable solution.

Algorithms based on the Suzuki-Trotter decomposition \cite{hdr,hdr1}
can solve (\ref{tdse}) for sufficiently long times (essential to determine
the pointer states of the central system).
They can handle Hamiltonians with explicit dependence on time,
are unconditionally stable, exactly preserve the unitarity
of quantum evolution, and the time step can be made
more than an order of magnitude bigger than in the
typical predictor-corrector method. Moreover, as our experience 
shows, for the scheme based on Suzuki-Trotter decomposition,
a large part of the total numerical error is accumulated in
the total phase of the wavefunction $|\Psi(t)\rangle$,
and does not affect any measurable physical quantities
(observables). However, for reasonably large systems, this scheme
is still slow, and simulations of decoherence lasted for
up to 200 CPU hours on a SGI 3800 supercomputer.
The problem of long simulation times becomes
especially prominent if we need to find the pointer states, or if the dynamics
of the central system is much faster than the decoherence rate.
We found that in these cicumstances, the method based on
Chebyshev's expansion becomes a very efficient tool to study problems of decoherence.
%
% new insert

Along with the Chebyshev's expansion method, the short iterative
Lanczos (SIL) approach \cite{LEFOR,sil,jaklic}, which is also 
based on the power-series expansion of the evolution operator, 
was found to be efficient for many similar problems of quantum
chemistry. We have tested this method, but our results
are negative. 
Low-order SIL method (with small number
of Lanczos iterations per step, usually,
less than 25) gives an unacceptable error, even for very short
time steps. On the other hand,
high-order SIL method (with more than 25 Lanczos iterations
per step) is
noticeably slower the approach based on Chebyshev's
expansion. 

We believe that low performance of SIL method originates from the
fact that for a small number of Lanczos
iterations (i.e., for low-order SIL), only a very
limited part of the spectrum is described correctly.
For a typical problem where SIL is known to be very effective
(e.g., a wavepacket propagation), most of relevant basis states 
have energy close to the energy of a wavepacket. 
Only these relevant states should be accurately described, while
accurate description of the whole energy spectrum is excessive. 
In contrast, in a typical spin-bath decoherence problem, 
a large number of bath states with very different energies
are involved in the decoherence process. Correspondingly,
a large part of spectrum should be taken into account,
and the high-order SIL integrator should be employed,
reducing the performance of the SIL method.

We also note that significant speed-up can be achieved
by using an approximate form of the wave function of the total
system (central system plus bath). In particular, the
multi-configurational time-dependent Hartree (MCTDH) method
\cite{mctdh1,mctdh2,mctdh3} is known to be very efficient, e.g.\ for
modeling of boson-bath decoherence.
The MCTDH approach uses an
approximate representation of the wave function, based on the
assumption that the wave function of the total system 
can be written as a superposition of a
relatively small number of "configurations", i.e. products of
time-varying single-spin 
wavefunctions. 

MCTDH is a method of choice
when the dimensionality of a single-particle Hilbert space is
large, and the multi-particle quantum correlations are associated
with a superposition of a small number of products of 
single-particle wavefunctions.
The problems considered in our paper present an opposite
situation. The bath consists of many spins 1/2, i.e. we have only
2 orbitals per particle (spin), and the single-particle evolution
is very simple, while the complex many-particle quantum
correlations are responsible for most of the physical effects
(i.e., the number of important single-spin-wavefunctions products
is very large). It is probable that many
problems of spin-bath decoherence can be efficiently treated
by MCTDH, but corresponding study requires a separate extensive 
research effort, which is beyond the framework of our paper.

% end new insert
%

\section{Chebyshev's method for spin-bath decoherence}
\label{sec3}

For a time-independent Hamiltonian, the
solution of Eq.\ (\ref{tdse}) can be formally written as
\begin{equation}
\Psi(t) = \exp{(-it{\cal H})} \Psi_0 = U(t) \Psi_0
\label{evop}
\end{equation}
where $U(t)=\exp{(-it{\cal H})}$ is the evolution operator.
An effective way \cite{TAL-EZER0,TAL-EZER,LEFOR,kosloff1,Iitaka01,SILVER,LOH}
of calculation of the exponent of a
large matrix ${\cal H}$ is to expand it in a series of
the Chebyshev polynomials of the operator ${\cal H}$.
Below, we describe the specific details of application of 
the Chebyshev method to the spin-bath decoherence simulations.

The Chebyshev's polynomials $T_k(x) = \cos{(k \arccos{x})}$
are defined for $x\in[-1,1]$. Thus, the Hamiltonian 
${\cal H}$ first should be rescaled
by the factor $E_0$ (the range of the values
of the system's energy) and shifted by $E_c$ (median value of the
systems' energy): 
\begin{eqnarray} 
\nonumber
E_c &=& \frac{1}{2}(E_{max}+E_{min}),\quad E_0 = E_{max}-E_{min}\\
\nonumber
E_{min} &=& \min\langle{\cal H}\rangle = 
  \min_{\langle\Phi|\Phi\rangle=1} \langle\Phi|{\cal H}|\Phi\rangle,\\
E_{max} &=& \max\langle{\cal H}\rangle =
  \max_{\langle\Phi|\Phi\rangle=1} \langle\Phi|{\cal H}|\Phi\rangle.
\end{eqnarray}
In this way, the rescaled operator ${\cal G}=
2({\cal H}-E_c)/E_0$ is also bounded by $-1$ and $1$:
$-1\le \langle{\cal G}\rangle \le 1$, i.e.\
$-1\le \langle\Phi|{\cal G}|\Phi\rangle\le 1$
for any state vector $|\Phi\rangle$ such that 
$\langle\Phi|\Phi\rangle=1$. For spin systems, the
Hamiltonian is bounded both from above and from below, 
and the operator ${\cal G}$ can be found.

In the specific case considered in this paper, when
the Hamiltonian ${\cal H}$ is defined by Eq.\ (\ref{specham}), 
we take
$E_0= 2\max{(|E_{min}|, |E_{max}|)}$. For this choice,
$-E_0/2 \le \langle{\cal H}\rangle\le E_0/2$. Correspondingly,
we can take $E_c=0$; this choice is legitimate, and, although
might be not optimal for some problems, still results in
very good performance of Chebyshev's method (see below).
Since $\max\langle{\cal H}\rangle=\|{\cal H}\|$ is the
norm of the Hamiltonian, the
value of $E_0$ can be estimated using the Cauchy's inequality:
$E_0/2\le \|{\cal H}_S\|+\|{\cal H}_B\|+\|{\cal V}\|$.
Similarly,
\begin{eqnarray}
\|{\cal H}_S\| &\le &
\sum_{\langle m,m'\rangle} \sum_{\alpha=x,y,z}
  |J^{\alpha}_{mm'}|\cdot
  \|S^{\alpha}_m\|\cdot \|S^{\alpha}_{m'}\| \\
  \nonumber
  &&+ \sum_m \sum_{\alpha=x,y,z}|H^\alpha_m|\cdot \|S^{\alpha}_m\|
  \\ \nonumber
  &=&
  \sum_{\langle m,m'\rangle} \sum_{\alpha=x,y,z}
  \frac 14 |J^{\alpha}_{mm'}|
   + \sum_m \sum_{\alpha=x,y,z}
  \frac 12 |H^\alpha_m|,
\end{eqnarray}
and $\|{\cal H}_B\|$ and $\|{\cal V}\|$ can be estimated in
the same manner. As a result, we have an estimate
$E_0\le E_1$, where
\begin{eqnarray}
\nonumber
E_1 &=& \sum_{\langle m,m'\rangle} \sum_{\alpha=x,y,z}
  \frac 12 |J^\alpha_{mm'}|
 + \sum_{\langle n,n'\rangle} \frac 12 |\Gamma^\alpha_{nn'}| \\
 && + \sum_{\langle m,n\rangle} \frac 12 |A^\alpha_{mn}|
 + \sum_m |H^\alpha_m| + \sum_n |H^\alpha_n|,
\end{eqnarray}
and the operator $\cal G$ can be defined as
${\cal G} = 2{\cal H}/E_1$, which satisfies the inequality
$-1\le \langle{\cal G}\rangle\le 1$.

The Chebyshev's expansion of the evolution operator $U(t)$ 
(see Eq.\ \ref{evop})
now looks like
\begin{equation}
U(t)=\exp{(-i\tau{\cal G})}=\sum_{k=0}^{\infty} c_k T_k({\cal G})
\label{chebinf}
\end{equation}
where $\tau=E_1 t/2$. The expansion
coefficients $c_k$
can be calculated using the orthogonal property
of the polynomials $T_k(x)$:
\begin{equation}
c_k= \frac {a_k}{\pi}  \int_{-1}^{1} \frac{T_k(x) \exp{(-ix\tau)}}
  {\sqrt{1-x^2}} \,dx = a_k (-i)^k J_k(\tau),
\end{equation}
where $J_k(\tau)$ is the Bessel function of $k$-th order,
and $a_k=2$ for $k=0$ and $a_k=1$ for $k\ge 1$. 
The successive terms in the Chebyshev's series can be efficiently
determined using the recursion 
\begin{equation}
T_{k+1}({\cal G}) = 2 {\cal G} T_k({\cal G}) + T_{k-1}({\cal G})
\label{recur}
\end{equation}
with the conditions $T_0({\cal G})=1$, $T_1({\cal G})={\cal G}$.
Thus, to find the vector $\Psi(t)$, we just need to sum
successively the terms of the series (\ref{chebinf}), using
Eq.\ (\ref{recur}) for calculation of the subsequent
terms, until we reach some pre-defined value $K$ of $k$,
which is determined by the required precision.

\begin{figure}[tbp!]
\includegraphics[width=8cm]{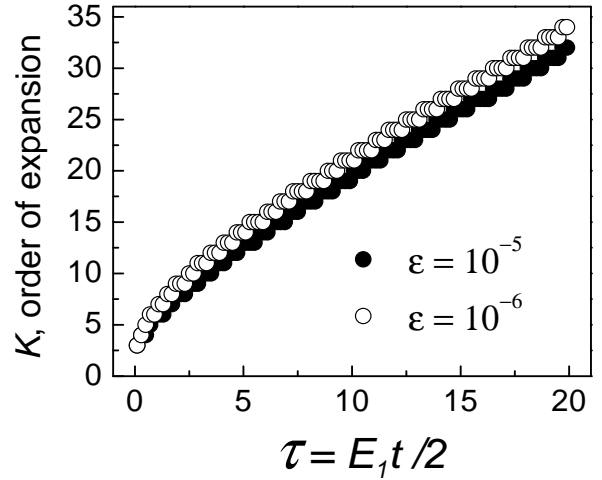}
\caption{Dependence of the order of the Chebyshev's expansion
$K$ on the value of $\tau=E_1t/2$. The solid circles corresponds
to the minimum value $\epsilon=10^{-5}$ of the expansion coefficient
$c_k$; the open circles corresponds to $\epsilon=10^{-6}$.}
\label{fignord}
\end{figure}

The high precision of this scheme originates from the fact that, for $k\gg \tau$,
the value of a Bessel function decreases super-exponentially
$J_k(\tau)\sim (\tau/k)^k$, so that termination of the 
series (\ref{chebinf}) at $k=K$ leads to an error which
decreases super-exponentially with $K$.
In practice, $K=1.5\tau$ already gives
precision of 10$^{-7}$ or better in most cases. Due to the
same reason, this scheme is asymptotically more efficient
than any time-marching scheme. For given sufficiently small
error $\epsilon$, the number of operations $Nop$ needed for
finding the wavefunction at time $T$, i.e. $\Psi(T)$,
grows linearly with $T$ for the Chebyshev-based
scheme. For a marching scheme of order $r$ with the
time step $\Delta t$, the numerical
error is $\epsilon\sim (\Delta t)^r T$, so that for 
given $\epsilon$ and $T$, the number of operations needed is
$Nop=T/\Delta t\sim T^{1+1/r}$, growing super-linearly with 
increasing $T$. 
For very long-time simulations, and when very
high precision is necessary, the Chebyshev method is
more efficient than any time-marching scheme known to us.
However, in practice, a precision better than 0.5\%--1\% is
very rarely needed. Similarly, very long-time simulations
are rarely of interest: in most cases,
the simulations are interesting only until the dynamics
of the system exhibits some non-trivial behavior. Therefore,
in spite of its asymptotic efficiency, the Chebyshev method is
not always the best choice for real research, and its
efficiency should be studied in every separate case.

\section{Simulation results}
\label{sec4}

We assess the usefulness of the Chebyshev's method for a wide
spectrum of decoherence problems, by consideriong two central problems
of decoherence, description of damping
of quantum oscillations in a system, and determination
of the pointer states.
In fact, there is no strict boundary: studying both 
problems, we track evolution of the system checking its
state at regular intervals of length $T$, but in studying
the oscillations decay the interval $T$ is much smaller than
the characteristic decoherence time $T_{dec}$, while 
in studying the pointer states, $T$ is larger than $T_{dec}$.

In spite of the asymptotic advantages of the Chebyshev-based
scheme, it is not {\it a priori\/} clear if it is
efficient for realistic problems, when the 
required numerical error $\delta$ is modest (say, 
$\delta=10^{-2}$--$10^{-3}$).
Also, if we track the dynamics of the decoherence
process, we make many steps of modest length $T$, and
the overhead associated with
the use of the Chebyshev's expansion might be significant,
see Fig.\ \ref{fignord}. 

To study this issue, we have performed several types of numerical tests.
The timing information reported in this paper has been
obtained from calculations on
a SGI Origin 3800 (500 MHz) system, using sequential, single processor code.
The order of Chebyshev's expansion $K$ have been defined
by the pre-specified precision $\epsilon$. We determined the 
minimum value of $K$ such that 
$|c_k|<\epsilon$ for $k\ge K$, starting from the
value $K_0=[1.1\tau]$ ($[x]$ is the integer part of $x$),
and adjusting it as needed.
Each simulation has been performed three times: (i)
using the Chebyshev's method with $\epsilon=10^{-12}$,
the reference run, (ii) using Chebyshev's method with
$\epsilon=10^{-5}$--10$^{-6}$, and (iii) using the
scheme based on Suzuki-Trotter decomposition \cite{hdr,hdr1}.
Previously we have used the latter to study spin-bath decoherence \cite{kondo,osc}.
In this paper, we have chosen to consider the same problems as in our
previous works on this subject, in order to avoid the impression that the tests have been
constructed to favor one particular method.

First, we consider the problem of oscillations damping
in the central system of two spins coupled by Heisenberg
exchange, interacting with the bath. We studied this
problem using the Suzuki-Trotter scheme in 
Ref.\ \onlinecite{osc}.
The Hamiltonians describing the bath and the
system are:
\begin{equation}
{\cal H}_S = J{\bf S}_1{\bf S}_2,\quad {\cal H}_B=0,\quad
{\cal V} = \sum A_n ({\bf S}_1 + {\bf S}_2) {\bf I}_n,
\label{hamstat}
\end{equation}
with $N=16$ bath spins. The exchange parameter $J=16.0$
(antiferromagnetic coupling between the central spins),
while $A_n$ are uniformly distributed between 0 and $-0.5$.
The initial state of the compound system 
$|\Psi_0\rangle=|\psi_0\rangle\otimes |\chi_0\rangle$
is the product of the initial state $|\psi_0\rangle$
of the central system, and $|\chi_0\rangle$ of the
bath. In this case, $|\psi_0\rangle=|\uparrow\downarrow\rangle$,
i.e.\ the first central spin is in the state $S^z_1(t=0)=+1/2$,
and the second spin is in the state $S^z_2(t=0)=-1/2$. The 
initial state of the bath $|\chi_0\rangle$ is the
linear superposition of all basis states with random 
coefficients. Physically, this situation corresponds to
the case of the temperature $\theta$ which is high
in comparison with the bath energies $A_n$, but is much lower
than the system's energy $J$ (note that $J\gg A_n$ in
this case).

\begin{figure}[tbp!]
\includegraphics[width=8cm]{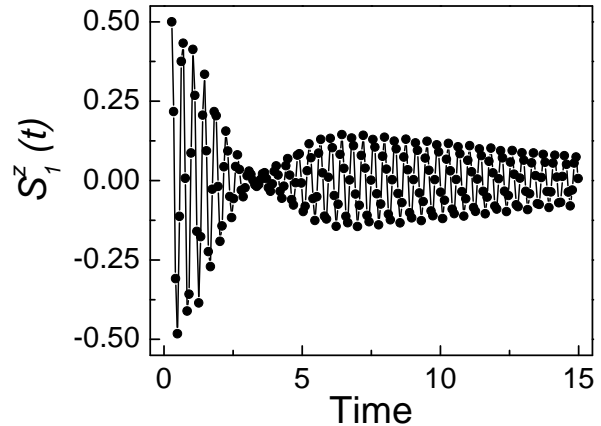}
\caption{Time dependence of the oscillations of the expectation value
of $S_1^z (t)$ in the two-spin system decohered by a spin bath.}
\label{figosc}
\end{figure}

The initial state of the central system is a
superposition of two eigenstates of ${\cal H}$: 
the state with 
the total spin $S=1$ and $S_z=0$, and the state
with the total spin $S=0$. These states have different
energies, and, for example, the dynamics of $S^z_1(t)$
is represented by oscillations with the frequency $J$.
Due to interaction with the spin bath, these oscillations
are damped, see Fig.\ \ref{figosc}. To study this damping
in detail, we take the Suzuki-Trotter
time step $\Delta t=0.035$, $T=2\Delta t$, and watch
the system since $t=0$ till $t_{max}=800 T$. If we do not
need such a high resolution, we increase $T$. In 
Table \ref{tabc}, we present the CPU time needed
to perform the simulations using the Suzuki-Trotter
and Chebyshev's methods, along with the resulting
error $\delta$ (which should not be confused with
the ``nominal'' precision of the Chebyshev's scheme
$\epsilon$). The error $\delta$ has been obtained
from comparison with the ``reference'' Chebyshev's 
run ($\epsilon=10^{-12}$), and is equal to the
maximum of absolute errors of the quantities
(all normalized to unity)
$2S^{\alpha}_1$, $2S^{\alpha}_2$,
$4S^{\alpha}_1S^{\beta}_1$  ($\alpha,\beta=x,y,z$), 
and the so-called ``quadratic entropy'' \cite{quent}
$S^{(2)}=1-\mathop{\rm Tr}\rho^2_S$. These quantities
have been calculated and compared at regular intervals 
of length $T$. Their calculation 
increases the number of computations, so that 
the tests 1, 2, and 3, which are otherwise equivalent 
for the Suzuki-Trotter method, require more and more
CPU time.

\begin{table}[bp!]
\caption{Comparison of the Suzuki-Trotter scheme (abbreviated
as ST) with the Chebyshev's scheme (abbreviated as Ch) for
the problem of oscillations decay.}
\label{tabc}
\begin{ruledtabular}
\begin{tabular}{ccccccc}
Test & $\Delta t$ & $T$ & $t_{max}$ & $\delta$ & $\epsilon$ & 
CPU Time \\
\hline
{\bf 1}, Ch & --- & $200\Delta t$ & $8 T$ & 
  $1\cdot 10^{-5}$ & $10^{-6}$ & 22 min \\
{\bf 1}, ST & 0.035 & $200\Delta t$ & $8 T$ & 
  $0.44\cdot 10^{-2}$ & --- & 80 min \\
\hline
{\bf 2}, Ch & --- & $8\Delta t$ & $200 T$ & 
  $0.3\cdot 10^{-4}$ & $10^{-6}$ & 59 min \\
{\bf 2}, ST & 0.035 & $8\Delta t$ & $200 T$ & 
  $0.48\cdot 10^{-2}$ & --- & 89 min \\
\hline
{\bf 3}, Ch & --- & $2\Delta t$ & $800 T$ & 
  $0.55\cdot 10^{-3}$ & $10^{-6}$ & 226 min \\
{\bf 3}, ST & 0.035 & $2\Delta t$ & $800 T$ & 
  $0.48\cdot 10^{-2}$ & --- & 156 min 
\end{tabular}
\end{ruledtabular}
\end{table}

As one can see from Table \ref{tabc}, for realistic
values of maximum error $\delta\sim 0.5\cdot 10^{-2}$,
and even for not very long runs, the Chebyshev's scheme
can be faster than the Suzuki-Trotter method by a factor of
up to four, and the efficiency of the Chebyshev's scheme grows
fast with increasing $T$. 
However, this straightforward comparison is too crude, and Table \ref{tabc} is only
an illustration of basic features of the Chebyshev's
method. To model fast oscillations which decay slowly
(often, with the decay time of order of decoherence time $T_{dec}$),
we should make $T$ significantly smaller than the 
oscillation period $t_{osc}=2\pi/J$, in order to 
correctly determine the amplitude of oscillations
at given time.

Therefore, to track the damping of oscillations, 
we use the two-leap approach: first, we make a
large time leap of length $T_1$ ($T_1\gg t_{osc}$,
but $T_1\ll T_{dec}$), and then we make a number $n_2$
(usually, 15--20) of smaller steps $T_2$ such that
$T_2\ll t_{osc}$ but $n_2 T_2\ge t_{osc}$, resolving
in detail one period of oscillations and extracting
the amplitude. By repeating this two-stage sequence
$n_{tot}$ times,
we can reliably track the change of the oscillations
amplitude with time.
The test example of this approach have been taken
from our recent work \cite{akakii}. We have performed
the same kind of simulations as described above,
with $N=16$ bath spins, repeating the two-leap
sequence $n_{tot}=8$ times, each time making one 
long leap $T_1$ followed by $n_2=21$ short leaps
$T_2$. The results of these tests are presented in
Table \ref{tabx}. Again, Chebyshev-based method
can be up to three times faster than the
Suzuki-Trotter algorithm \cite{hdr,hdr1}.

\begin{table}[tbp!]
\caption{Comparison of the Suzuki-Trotter scheme (abbreviated
as ST) with the Chebyshev's scheme (abbreviated as Ch) for
the problem of oscillations decay, employing the two-leap
approach with different $T_1$ and $T_2$.}
\label{tabx}
\begin{ruledtabular}
\begin{tabular}{ccccccc}
Test & $\Delta t$ & $T_1$ & $T_2$ & 
$\delta$ & $\epsilon$ & CPU Time \\
\hline
{\bf 4}, Ch & --- & $150\Delta t$ & $\Delta t$ & 
  $0.4\cdot 10^{-4}$ & $10^{-6}$ & 61 min \\
{\bf 4}, ST & 0.02 & $150\Delta t$ & $\Delta t$ & 
  $0.2\cdot 10^{-2}$ & --- & 144 min \\
\hline
{\bf 5}, Ch & --- & $300\Delta t$ & $\Delta t$ & 
  $0.4\cdot 10^{-4}$ & $10^{-6}$ & 75 min \\
{\bf 5}, ST & 0.02 & $300\Delta t$ & $\Delta t$ & 
  $0.3\cdot 10^{-2}$ & --- & 221 min 
\end{tabular}
\end{ruledtabular}
\end{table}

Finally, we have tested the Chebyshev scheme in the
problem of determining the pointer states, employing
an example from our work \cite{kondo}. This example is
interesting also because it deals with a physically
important case of a spin bath possessing chaotic
internal dynamics, which is relevant 
for majority of realistic spin baths (such as nuclear spins or 
impurity spins baths). 
The Hamiltonian describing the
system is 
\begin{equation}
{\cal H}_S = J{\bf S}_1{\bf S}_2,\quad 
{\cal V} = \sum A_n {\bf S}_1 {\bf I}_n,
\label{hamchaos0}
\end{equation}
i.e., the bath spins are coupled only with the first
central spin, and the bath Hamiltonian is now
\begin{equation}
{\cal H}_B=\sum_n h_z I^z_k + \sum_{\langle n,n'\rangle}
  U_{n n'} I^x_n I^x_{n'}.
\label{hamchaos1}
\end{equation}
In our simulations we used $h=0.1$ and
$U_{n n'}$ randomly distributed in the interval 
$[-0.013,0.013]$. This Hamiltonian is known to result
in stochastic behavior \cite{shepel}; we have
checked the level statistics independently, and 
found that it closely follows the Wigner-Dyson distribution.

To determine the pointer states, we need to find
the elements of the reduced
density matrix $\rho_S (t)$ in the long-time limit 
$t\to\infty$. We start at $t=0$ from the 
state of the compound system which is the product 
of the states of the bath and the central system (as
above), but the initial state of the central spins now
is the singlet 
$|\psi_0\rangle=(1/\sqrt{2})[|\uparrow\downarrow\rangle
-|\downarrow\uparrow\rangle]$. 
Because of decoherence, the final
state of the central system is mixed, and 
$\rho_S= w_1|p_1\rangle\langle p_1| + w_2|p_2\rangle\langle p_2|$,
where $|p_1\rangle$ and $|p_2\rangle$ are the pointer
states, which are 
superpositions of the states $|\uparrow\uparrow\rangle$, 
$|\downarrow\downarrow\rangle$
$|\uparrow\downarrow\rangle$,
and $|\downarrow\uparrow\rangle$. As we have found in
our work \cite{kondo}, the form of this 
superposition is determined by the ratio $J/b$, where
$b=\sum_n A^2_n$. For $J/b\sim 1$, the pointer states
are very close to the singlet $S=0$ and triplet
$S=1$, $S_z=0$ states, and for $J\ll b$, the pointer
states are close to $|\uparrow\downarrow\rangle$
and $|\downarrow\uparrow\rangle$. Thus, the 
quantities characterizing the type of the pointer state
are the values of the non-diagonal elements of the
density matrix $\rho_S$ in the basis 
$|\uparrow\uparrow\rangle$, $|\downarrow\downarrow\rangle$
$|\uparrow\downarrow\rangle$,
and $|\downarrow\uparrow\rangle$. In particular,
the element 
$\rho^{12}_S=\langle\uparrow\downarrow|\rho_S|
\downarrow\uparrow\rangle$ is a very suitable
quantity to characterize the pointer state.
This non-diagonal
element is close to zero for $J\ll b$, and gradually increases
in absolute value with increasing $J$.

\begin{figure}[tbp!]
\includegraphics[width=8cm]{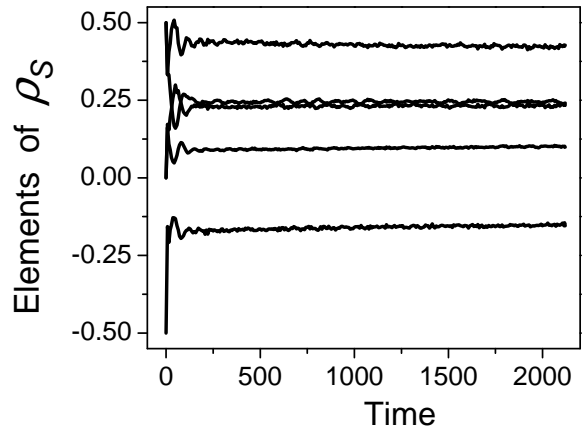}
\caption{Temporal evolution of different
elements of the density matrix $\rho$: diagonal elements
corresponding to the states $|\uparrow\uparrow\rangle$,
$|\uparrow\downarrow\rangle$, $|\downarrow\uparrow\rangle$,
and $|\downarrow\downarrow\rangle$
(the four upper curves),
and the non-diagonal element $\rho^{12}_S$ (the lowest curve).
Very slow relaxation is better seen for the uppermost
curve 
(the diagonal element corresponding to the state
%$\langle\uparrow\uparrow|\rho_S|\uparrow\uparrow\rangle$
$|\uparrow\uparrow\rangle$)
which has a small negative slope.
Note that the two lines in the middle (the second and the third 
lines from above, the diagonal elements corresponding to the
states
%$\langle\uparrow\downarrow|\rho_S|\uparrow\downarrow\rangle$
$|\uparrow\downarrow\rangle$
and
%$\langle\downarrow\uparrow|\rho_S|\downarrow\uparrow\rangle$
$|\downarrow\uparrow\rangle$,
correspondingly) are very close to each other 
at $t\ge 200$, as expected
for a near-equilibrium (although not completely
relaxed) situation.}
\label{figdm}
\end{figure}

Typical results for temporal evolution of the elements of the
density matrix $\rho_S$ are shown in Fig.\ \ref{figdm}.
One can see that in this situation,
we {\it do not\/} need to use the two-leap
approach with different $T_1$ and $T_2$.
The relaxation (after some initial period) is slow,
and no fast oscillations of considerable amplitude
exist at long times, so that the one-leap approach is sufficient.
Thus, the efficiency of the Chebyshev-based scheme
is expected to be very good.
This is indeed the case, as Table \ref{tabt} 
demonstrates. The results presented there correspond
to $J=0.1$. The Chebyshev-based scheme is faster
than the Suzuki-Trotter method up to a factor of 8.

\begin{table}[tbp!]
\caption{Comparison of the Suzuki-Trotter scheme (abbreviated
as ST) with the Chebyshev's scheme (abbreviated as Ch) for
the problem of determining the pointer states.}
\label{tabt}
\begin{ruledtabular}
\begin{tabular}{ccccccc}
Test & $\Delta t$ & $T$ & $t_{max}$ & $\delta$ & $\epsilon$ &
CPU Time \\
\hline
{\bf 6}, Ch & --- & $100\Delta t$ & $500 T$ &
  $0.2\cdot 10^{-3}$ & $10^{-6}$ & 19 min \\
{\bf 6}, ST & 0.14 & $100\Delta t$ & $500 T$ &
  $0.7\cdot 10^{-2}$ & --- & 105 min \\
\hline
{\bf 7}, Ch & --- & $10\Delta t$ & $5000 T$ &
  $0.1\cdot 10^{-2}$ & $10^{-6}$ & 52 min \\
{\bf 7}, ST & 0.14 & $10\Delta t$ & $5000 T$ &
  $0.8\cdot 10^{-2}$ & --- & 117 min \\
\hline
{\bf 8}, Ch & --- & $1000\Delta t$ & $50 T$ &
  $0.3\cdot 10^{-6}$ & $10^{-6}$ & 13 min \\
{\bf 8}, ST & 0.14 & $1000\Delta t$ & $50 T$ &
  $0.8\cdot 10^{-2}$ & --- & 107 min \\
\end{tabular}
\end{ruledtabular}
\end{table}

We have checked our conclusions on many other cases,
with the central systems made of up to $M=4$ spins,
and with the baths made of up to $N=22$ spins, with
different Hamiltonians and different 
values of the Hamiltonian parameters.
We found that Chebyshev-based method gives a significant
increase in the simulations speed for all 
problems where the value of $T$ can be made sufficiently large.

\section{Summary}
\label{secsum}

Theoretical studies of the spin-bath decoherence are important for many
areas of physics, including quantum mechanics and quantum measurement
theory, quantum computing, solid state physics etc. Decoherence is a 
complex many-body phenomenon, and numerical simulation is an
important tool for its investigation. In this paper, we have studied
efficiency of the numerical scheme based on the Chebyshev expansion.
We have presented specific details
of the application of this method
to the spin-bath decoherence modeling.
To assess the efficiency of the simulation method,
we have used model problems which we have encountered in our previous studies
of the spin-bath decoherence.
We compared the Chebyshev-based
scheme with a fast method based on the Suzuki-Trotter decomposition.
We have found that in many cases, the former gives a considerable
increase in the speed of simulations, sometimes up to a factor of
eight (for the problem of finding the system's pointer states),
while in studying the decoherence dynamics, the increase in
speed is less drastic (a factor of 2--3), but still considerable.
This conclusion holds for many types of central systems and
spin baths, with different Hamiltonians.
\begin{acknowledgments}
This work was partially carried
out at the Ames Laboratory, which is operated for the U.\ S.\ Department of
Energy by Iowa State University under Contract No.\ W-7405-82 and was
supported by the Director of the Office of Science, Office of Basic Energy
Research of the U.\ S.\ Department of Energy.
Support from the
Dutch ``Stichting Nationale Computer Faciliteiten (NCF)''
is gratefully acknowledged.
\end{acknowledgments}


\begin{thebibliography}{99}
\bibitem{quent} Quadratic entropy $S^{(2)}=1-\mathop{\rm Tr}\rho^2_S$
  characterizes, how mixed is the state of the central system; for pure 
  states $S^{(2)}=0$.
\end{thebibliography}
\end{document}